\documentclass[conference]{IEEEtran}
\IEEEoverridecommandlockouts
\usepackage{ucs}
\usepackage[utf8x]{inputenc}
\usepackage[cmex10]{amsmath}
\usepackage{cite, amsfonts, amssymb, amsthm, bm, bbm, graphicx, relsize, multirow, booktabs, color, colortbl,url,soul}
\usepackage[american]{babel}
\usepackage[T1]{fontenc}
\usepackage{CJK}
\usepackage{algorithmic, algorithm}
\usepackage{lettrine}
\usepackage{glossaries}
\usepackage{upgreek}
\usepackage{stfloats}

\usepackage{flushend}
\usepackage{subfigure}
\theoremstyle{remark}

\graphicspath{{figures/}}

\title{Subspace-Based Detection and Localization in Distributed MIMO Radars}
	
\author{\IEEEauthorblockN{Yangming~Lai\IEEEauthorrefmark{1}, 
		Luca~Venturino\IEEEauthorrefmark{2}\IEEEauthorrefmark{3},
		Emanuele~Grossi\IEEEauthorrefmark{2}\IEEEauthorrefmark{3},		
		Wei~Yi\IEEEauthorrefmark{1}}
	\IEEEauthorblockA{Email: ymlaiuestc@163.com, l.venturino@unicas.it, e.grossi@unicas.it, kussoyi@gmail.com }
	\IEEEauthorblockA{\IEEEauthorrefmark{1}School of Information and Communication Engineering, University of Electronic Science and Technology of China, China}
	\IEEEauthorblockA{\IEEEauthorrefmark{2}Department of Electrical and Information Engineering, University of Cassino and Southern Lazio,  Italy}
	\IEEEauthorblockA{\IEEEauthorrefmark{3}National Inter-University Consortium for Telecommunications, Italy}
   
   \thanks{This work was supported in part by the National Natural Science Foundation of China under Grants 61771110, U19B2017, and 61871103, and in part by the Fundamental Research Funds of Central Universities under Grant ZYGX2020ZB029.}
}

\newacronym{papr}{PAPR}{peak-to-average power ratio}
\newacronym{lfm}{LFM}{linear frequency modulated}
\newacronym{pn}{PN}{pseudo-noise}
\newacronym{pri}{PRI}{pulse repetition interval}
\newacronym{dmg}{DMG}{directional multi-gigabit}
\newacronym{dti}{DTI}{data transmission interval}
\newacronym{ap}{AP}{access point}
\newacronym{dfrc}{DFRC}{dual function radar communication}
\newacronym{mcss}{MCSs}{modulation and coding schemes}
\newacronym{cbaps}{CBAPs}{contention-based access periods}
\newacronym{ssps}{SSPs}{scheduled service periods}
\newacronym{cbap}{CBAP}{contention-based access period}
\newacronym{sp}{SP}{scheduled service period}
\newacronym{stf}{STF}{short training field}
\newacronym{cef}{CEF}{channel estimation field}

\newacronym{aic}{AIC}{Akaike information criterion}
\newacronym{bic}{BIC}{Bayesian information criterion}
\newacronym{gic}{GIC}{generalized information criterion}

\newacronym{stb}{STB}{single-target bound}
\newacronym{cfar}{CFAR}{constant false alarm rate}
\newacronym{far}{FAR}{false alarm rate}
\newacronym{rcs}{RCS}{radar cross-section}
\newacronym{amf}{AMF}{adaptive matched filter}
\newacronym{mf}{MF}{matched filter}
\newacronym{mf-pd}{MF-PD}{matched-filter peak detector}
\newacronym{wmf}{WMF}{whitening matched filter}
\newacronym{br}{BR}{beam refinement}
\newacronym{iss}{ISS}{initiator sector sweep}
\newacronym{rss}{RSS}{responder sector sweep}
\newacronym{ssw-fb}{SSF}{sector sweep feedback}
\newacronym{ssw-ack}{SSA}{sector sweep acknowledgement}
\newacronym{iaa}{IAA}{iterative adaptive approach}
\newacronym{iaa-apes}{IAA-APES}{iterative adaptive approach for amplitude and phase estimation}
\newacronym{iic}{IIC}{iterative interference cancellation}
\newacronym{bpsk}{BPSK}{binary phase shift keying}

\newacronym{mac}{MAC}{multiple-access channel}
\newacronym{bc}{BC}{broadcast channel}
\newacronym{mimo}{MIMO}{multiple-input multiple-output}
\newacronym{siso}{SISO}{single-input single-output}
\newacronym{af}{AF}{amplify-and-forward}
\newacronym{df}{DF}{decode-and-forward}
\newacronym{cf}{CF}{compress-and-forward}
\newacronym{mwrc}{MWRC}{multi-way relay channel}
\newacronym{dmmwrc}{DM-MWRC}{discrete memoryless multi-way relay channel}
\newacronym{pde}{PDE}{partial data exchange}
\newacronym{fde}{FDE}{full data exchange}
\newacronym{iid}{i.i.d.}{independent and identically distributed}
\newacronym{awgn}{AWGN}{additive white Gaussian noise}
\newacronym{awg}{AWG}{additive white Gaussian}

\newacronym{snr}{SNR}{signal-to-noise ratio}
\newacronym{sinr}{SINR}{signal-to-interference-plus-noise ratio}
\newacronym{inr}{INR}{interference to noise ratio}
\newacronym{zf}{ZF}{zero-forcing}
\newacronym{mmse}{MMSE}{minimum mean square error}
\newacronym{sud}{SUD}{single user decoding}
\newacronym{dof}{DoF}{degrees of freedom}
\newacronym{gdof}{GDoF}{generalized degrees of freedom}
\newacronym{nnc}{NNC}{noisy network coding}
\newacronym{dmn}{DMN}{discrete memoryless network}
\newacronym{csi}{CSI}{channel state information}
\newacronym{pmf}{pmf}{probability mass function}
\newacronym{dmic}{DM-IC}{discrete memoryless interference channel}
\newacronym{ic}{IC}{interference channel}
\newacronym{ee}{EE}{energy efficiency}
\newacronym{gee}{GEE}{global energy efficiency}
\newacronym{ian}{IAN}{treating interference as noise}
\newacronym{snd}{SND}{simultaneous non-unique decoding}
\newacronym{hk}{HK}{Han-Kobayashi}
\newacronym{rf}{RF}{radio frequency}
\newacronym{pa}{PA}{power amplifier}
\newacronym{lna}{LNA}{low noise amplifier}
\newacronym{lo}{LO}{local oscillator}
\newacronym{adc}{ADC}{analog-to-digital converter}
\newacronym{dac}{DAC}{digital-to-analog converter}
\newacronym{dsp}{DSP}{digital signal processing}
\newacronym{brd}{BRD}{best response dynamics}
\newacronym{ne}{NE}{Nash equilibrium}
\newacronym{lhs}{LHS}{left-hand side}
\newacronym{rhs}{RHS}{right-hand side}
\newacronym{glrt}{GLRT}{generalized likelihood ratio test}
\newacronym{lr}{LR}{likelihood ratio}
\newacronym{llr}{LLR}{log-LR}
\newacronym{ml}{ML}{maximum likelihood}
\newacronym{mle}{MLE}{maximum likelihood estimation}
\newacronym{los}{LOS}{line-of-sight}
\newacronym{sls}{SLS}{sector level sweep}
\newacronym{cphy}{CPHY}{control physical}
\newacronym{scphy}{SCPHY}{single-carrier physical}
\newacronym{qo}{QO}{quasi omni-directional}
\newacronym{ssw}{SSW}{sector sweep}
\newacronym{nrmse}{NRMSE}{normalized root mean square error}
\newacronym{crb}{CRB}{Cram\'er-Rao bound}
\newacronym{cdf}{CDF}{cumulative distribution function}
\newacronym{pdf}{PDF}{probability density function}
\newacronym{mvdr}{MVDR}{minimum variance distortionless response}
\newacronym{pfa}{Pfa}{probability of false alarm}
\newacronym{pd}{Pd}{probability of detection}
\newacronym{ssr}{SSR}{successive space removal }
\newacronym{sic}{SIC}{successive interference cancellation}
\newacronym{doa}{DOA}{direction-of-arrival}

\newacronym{svd}{SVD}{singular value decomposition}
\newacronym{cdma}{CDMA}{code division multiple access}
\newacronym{tdma}{TDMA}{time-division multiple access}
\newacronym{fdma}{FDMA}{frequency-division multiple access}
\newacronym{ddma}{DDMA}{doppler division multiple access}
\newacronym{msdis}{MSD-IS}{matched subspace detector with iterative estimation of the interference subspace}
\newacronym{msdicm}{MSD-ICM}{matched subspace detector with iterative estimation of the interference covariance matrix}
\newacronym{toa}{TOA}{time of arrival}
\newacronym{aoa}{AOA}{angle of arrival}
\newacronym{crlb}{CRLB}{Cramer-Rao lower bound}
\newacronym{rmse}{RMSE}{root mean square error}

\newacronym{jmdlsic}{JMDL-SIC}{joint multi-target detection and localization based successive interference cancellation}

\newacronym{jdl}{JDL}{joint detection and localization}

\newacronym{mc}{MC}{Monte Carlo}
\newacronym{glrtcd}{GLRT-CD}{generalized likelihood ratio test with cleaned data}

\newacronym{jdlsic}{JDL-SIC}{JDL algorithm with successive interference cancellation}
\newacronym{jdlssr}{JDL-SSR}{JDL based successive space removal}

\begin{document}
\bstctlcite{BSTcontrol}

\maketitle

\begin{abstract}
	In this paper, we consider a distributed multiple-input multiple-output (MIMO) radar which radiates waveforms with non-ideal cross- and auto-correlation functions and derive a novel subspace-based procedure to detect and localize multiple prospective targets. The proposed solution solves a sequence of composite binary hypothesis testing problems by resorting to the generalized information criterion (GIC); in particular, at each step, it aims to detect and localize one additional target, upon removing the interference caused by the previously-detected targets. An illustrative example is provided.	
\end{abstract}
\begin{IEEEkeywords}
	Detection and localization, distributed MIMO radar, generalized information criterion, non-ideal correlation.
\end{IEEEkeywords}

\section{Introduction}\label{sect:Introduction}

Distributed \gls{mimo} radars are equipped with widely-spaced transmitters and receivers that usually see different aspect angles of a prospective target~\cite{fishler2006spatial,haimovich2007mimo,wang2011moving,hassanien2012moving}. There are two main methodologies to perform detection and localization in these radar architectures. On the one hand, each receiver first elaborates its own data to identify candidate detections, while a fusion center makes a final decision on the target presence and performs its localization via triangulation~\cite{dianat2013target,park2015closed1,amiri2017exact}. On the other hand, the raw signals of all receivers are jointly elaborated to accomplish the desired task~\cite{godrich2010target,he2010noncoherent,bar2011direct,Yi-2020}: this latter strategy will be considered next. 

The works in~\cite{godrich2010target,he2010noncoherent,bar2011direct} have discussed localization algorithms which rely upon the maximum likelihood estimation of the unknown parameters;  however, these solutions do not consider the detection task and cannot handle an unknown target number. \Gls{jdl} is instead considered in~\cite{Yi-2020}; however, this study assumes that the radiated waveforms have an all-zero cross-correlation and a thumb-tack auto-correlation function, so that a \gls{mf} based receiver can be employed. However, such ideal correlation properties cannot be obtained in practice, and the resulting sidelobes can significantly degrade the  performance~\cite{ma2010receiver,Abramovich-2008,zhao2020suppression,wang2020target,li2021signal,zeng2021delay}, if not properly accounted for in the design. 

In this study, we consider the \gls{jdl} of multiple targets in a distributed \gls{mimo} radar with an imperfect waveform separation.  Interestingly, a somehow similar problem has been recently considered in the context of mmWave dual-function radar-communication systems~\cite{GLV-CAMSAP2019,grossi2020adaptive,GLV-Radar2020,GLV-2021-TWC} and of passive radars~\cite{Colone-2016,Blasone-2020}. Inspired by these related studies, we make the following contributions. We  show that the superposition of the echoes produced by each target in response to waveforms emitted by the radar transmitters can be regarded as a subspace signal, whose structure is specified by the target location. Following the design methodology in~\cite{GLV-2021-TWC}, we derive a novel iterative subspace-based detector that extracts one target at the time from the data and is robust to the sidelobe masking caused by the stronger targets on weaker ones. The proposed procedure generalizes the one discussed  in~\cite{GLV-2021-TWC} to the case where multiple widely-spaced transmitters/receivers are present and the targets belong to linear subspaces which may be not independent. Finally, we assess the performance of the proposed detector, also in comparison with the solution in~\cite{Yi-2020} and the single-target benchmark. 

The remainder of this work is organized as follows. Sec.~\ref{sect:System and Signal models} contains the signal model. Sec.~\ref{sect: Proposed Strategies_total} presents the proposed algorithm. Sec.~\ref{sect:Simulation example} provides  the numerical analysis. Finally, the conclusions are given in Sec.~\ref{sect:Conclusion and next works}.

\section{Signal Model}\label{sect:System and Signal models}

Consider a distributed MIMO radar with $N$ transmitters located at $\{\bm{x}^{\rm tx}_{n}\}_{n=1}^{N}$ and $P$ receivers located at $\{\bm{x}^{\rm rx}_{p}\}_{p=1}^{P}$, with $\bm{x}^{\rm tx}_{n},\bm{x}^{\rm rx}_{p}\in\mathbb{R}^{2}$; the $n$-th transmitter emits the signal $\tilde{s}_{n}(t)$ which has a two-sided bandwidth $W$ and supports in $[0,T]$, with $W T\gg N$. Let $f_{\max}\geq0$ be the maximum magnitude of the Doppler shift in any target echo, we assume $f_{\max}T\ll 1$, so that such Doppler shift can be neglected. If $K\in\{0,1,\ldots,K_{\max}\}$ point-like targets are present, the baseband signal observed by the $p$-th receiver is~\cite{godrich2010target}
\begin{equation}\label{r_pt_tilde}
\tilde{r}_{p}(t)=  \sum_{k=1}^{K} \sum_{n=1}^{N} a_{p,n,k}\tilde{s}_{n}\big(t-\tau_{p,n}(\bm{x}_{k})\big)+\tilde{w}_{p}(t)
\end{equation}
for $p=1,\ldots,P$. Here, $\tilde{w}_{p}(t)$ is the additive noise, modeled as a complex circularly-symmetric Gaussian process independent across the receivers. Also, the first term on the right hand side is present only if $K\geq 1$; in this case, $\bm{x}_{k}\in\mathbb{R}^{2}$ is the location of the $k$-th target, while $a_{p,n,k}\in\mathbb{C}$ and $\tau_{p,n}(\bm{x}_{k})=  \|\bm{x}^{\rm rx}_{p}-\bm{x}_{k}\|/c+\|\bm{x}^{\rm tx}_{n}-\bm{x}_{k}\|/c$ are the amplitude and the delay of its echo with respect to the receiver/transmitter pair $(p,n)$, respectively, and $c$ is the speed of light. The number of targets and their amplitudes and locations are unknown.

To remove the out-of-bandwidth noise, $\tilde{r}_{p}(t)$ is sent to a low-pass filter, whose impulse response $\phi(t)$ has support in $[0, T_{\phi}]$. Upon defining $r_{p}(t)=\tilde{r}_{p}(t)\star \phi(t)$, $s_{n}(t)=\tilde{s}_{n}(t)\star \phi(t)$, and $w_{p}(t)=\tilde{w}_{p}(t)\star \phi(t)$, we have
\begin{equation}\label{r_pt}
r_{p}(t)=  \sum_{k=1}^{K} \sum_{n=1}^{N} a_{p,n,k}s_{n}\big(t-\tau_{p,n}(\bm{x}_{k})\big)+w_{p}(t).
\end{equation}
We make here the standard assumption that two copies of $s_{n}(t)$ with delay $\delta_{1}$ and $\delta_{2}$ are resolvable if $|\delta_{1}-\delta_{2}|>1/W$, and, following~\cite{Yi-2020}, we say that $K\geq 2$ targets located at $\{\bm{x}_{k}\}_{k=1}^{K}$ are \emph{separable} if, for any target pair $(k,q)$, with $k\neq q$, there exists at least one receiver/transmitter pair $(p,n)$ such that $s_{n}\big(t-\tau_{p,n}(\bm{x}_{k})\big)$ and $s_{n}\big(t-\tau_{p,n}(\bm{x}_{q})\big)$ are resolvable,

Finally, $r_{p}(t)$ is sampled at rate $1/T_{s}$ in the interval  $\mathcal{I}=[\tau_{\min},T+T_{\phi}+\tau_{\max})$, where $\tau_{\min}$ and $\tau_{\max}$ are the minimum and maximum delays of a prospective echo, respectively; the  $M=\left \lceil (T+T_{\phi}+\tau_{\max}-\tau_{\min})/T_{s} \right \rceil$ data samples are organized into the following $M$-dimensional vector
\begin{equation}\label{eq:discrete_signal_per_antenna}
	\bm{r}_{p}=\sum_{k=1}^{K} \bm{S}_{p}(\bm{x}_{k}) \bm{a}_{p,k}+\bm{w}_{p}
\end{equation}
where $\bm{S}_{p}(\bm{x}_{k})=\left[\bm{s}_{p,1}(\bm{x}_{k})\,\cdots\,\bm{s}_{p,N}(\bm{x}_{k})\right]\in\mathbb{C}^{M \times N}$, $\bm{s}_{p,n}(\bm{x}_{k})$  contains the samples $\big\{s_{n}\big((m-1)T_{s}+\tau_{\min}-\tau_{p,n}(\bm{x}_{k})\big)\big\}_{m=1}^{M}$ and represents the signature of the echo generated by the $k$-th target towards the $p$-th receiver when illuminated by the $n$-th transmitter,  $\bm{a}_{p,k}=(a_{p,1,k},\ldots, a_{p,N,k})^{\top}\in\mathbb{C}^{N}$, and $\bm{w}_{p}\in\mathbb{C}^{M}$ is a complex circularly-symmetric Gaussian vector with full-rank covariance matrix $\bm{C}_{p}$.

\section{Proposed Subspace-Based Detector}\label{sect: Proposed Strategies_total}
We aim to jointly detect and localize multiple targets, given only the location of the transmitters and receivers, the emitted waveforms, and the measurement vectors $\bm{r}_{1},\ldots,\bm{r}_{P}$. To proceed, notice that $\bm{r}_{p}$ is the noisy superposition of an unknown number of \emph{subspace signals} originated from as many targets; the $k$-th subspace signal belongs to the column span of the mode matrix  $\bm{S}_{p}(\bm{x}_{k})$, which only depends upon the  target location $\bm{x}_{k}$. Leveraging the design methodology in~\cite{GLV-2021-TWC}, we propose to solve a sequence of composite binary hypothesis testing problems: in each problem, we aim to detect a subspace signal generated by a prospective target with an unknown location in the presence of the subspace interference caused by the previously-detected targets plus independent noise. 

As customary, we start by assuming that the targets are located on a finite grid, say $\mathcal{G}$, whose points are uniformly-spaced in the inspected region and have an inter-element spacing $\Delta_{g} \leq c/W$; also, we restrict the search to targets which are at least separable. Let $\hat{\bm{x}}^{(k)}$ be the estimated location of the target detected in the $k$-th iteration; also, let $\mathcal{G}^{(k)}$ be the search set in the $k$-th iteration, with $\mathcal{G}^{(1)}=\mathcal{G}$ and
\begin{multline}\label{grid_update}
	\mathcal{G}^{(k)}=\Big\{
	\bm{g}\in\mathcal{G}^{(k-1)}:\\
	\min_{i\in\{1,\ldots,k-1\}}\max_{\substack{p\in\{1,\ldots,P\} \\ n\in\{1,\ldots,N\}}}\;\big|\tau_{p,n}(\bm{g})-\tau_{p,n}(\hat{\bm{x}}^{(i)})\big|>\frac{1}{W}\Big\}
\end{multline}
for $k\geq2$. Then, the $k$-th testing problem to be solved is
\begin{equation}\label{bar_H_k}
	\begin{cases}
		\mathcal{H}^{(k)}_{0}: & \displaystyle \bm{r}_{p}=\sum_{i=1}^{k-1}\bm{S}_{p}(\hat{\bm{x}}^{(i)})\bm{a}_{p}^{(i)} + \bm{w}_{p}, \; \forall \;p\\[5pt]
		\mathcal{H}^{(k)}_{1}: & \displaystyle\bm{r}_{p}= \bm{S}_{p}(\bm{x}^{(k)}) \bm{a}_{p}^{(k)}\\
		& \displaystyle \hspace{0.8cm}+\sum_{i=1}^{k-1}\bm{S}_{p}(\hat{\bm{x}}^{(i)})\bm{a}_{p}^{(i)} + \bm{w}_{p}, \;\forall \;p	
	\end{cases}
\end{equation}
for $k=1,2,\ldots,K_{\max}$. At the $p$-th receiver,  $\bm{S}_{p}(\bm{x}^{(k)})\in\mathbb{C}^{M\times N}$ is the mode matrix of a prospective target located in $\bm{x}^{(k)}\in \mathcal{G}^{(k)}$,  while $\bm{a}_{p}^{(k)}=\text{cat}\big\{a_{p,1}^{(k)},\ldots,a_{p,N}^{(k)}\big\}\in\mathbb{C}^{N}$ is the corresponding unknown gain vector; also, for $k\geq 2$, $\bm{S}_{p}(\hat{\bm{x}}^{(i)})\in\mathbb{C}^{M\times N}$ is the mode matrix of the interference  caused by the target detected in the $i$-th iteration, while $\bm{a}_p^{(i)} \in\mathbb{C}^{N}$ is the corresponding unknown gain vector, for $i=1,\ldots,k-1$. The negative log-likelihood functions under $\mathcal{H}_{0}^{(k)}$ and $\mathcal{H}_{1}^{(k)}$ at the $p$-th receiver are
\begin{multline}\label{nega_log_H_0_k}
	-\ln {{f}_{p,0}^{(k)}}\big( {\bm{r}_{p}};\bm{\alpha}_{p,0}^{(k)} \big)=
	\ln \Big(\pi ^{M}\det  \bm{C}_{p}  \Big)\\ +\Big\| \bm{C}_{p}^{-1/2}\Big( {\bm{r}_{p}}-\bm{\Sigma}_{p,0}^{(k)}\bm{\alpha}_{p,0}^{(k)} \Big)\Big\|^2
\end{multline}
and
\begin{multline}\label{nega_log_H_1_k}
	-\ln {{f}_{p,1}^{(k)}}\big( {\bm{r}_{p}};\bm{x}^{(k)},\bm{\alpha}_{p,1}^{(k)}\big)=	\ln \Big(\pi ^{M}\det \bm{C}_{p} \Big) \\+\Big\| \bm{C}_{p}^{-1/2}\Big( {\bm{r}_{p}}-\bm{\Sigma}_{p,1}^{(k)}(\bm{x}^{(k)}) \bm{\alpha}_{p,1}^{(k)}\Big)\Big\|^2
\end{multline}
respectively, where
\begin{subequations}
	\begin{align}
		\bm{\Sigma}_{p,0}^{(k)}&=\begin{bmatrix}\bm{S}_{p}(\hat{\bm{x}}^{(1)}) &  \cdots & \bm{S}_{p}(\hat{\bm{x}}^{(k-1)})\end{bmatrix}\in\mathbb{C}^{M\times (k-1)N}\\
		\bm{\alpha}_{p,0}^{(k)}&=\text{cat}\big\{\bm{a}_{p}^{(1)},\ldots,\bm{a}_{p}^{(k-1)}\big\}\in\mathbb{C}^{(k-1)N}
	\end{align}
\end{subequations}
are the augmented mode matrix and gain vector of the interference under  $\mathcal{H}_{0}^{(k)}$, while
\begin{subequations}
	\begin{align}	
		\bm{\Sigma}_{p,1}^{(k)}(\bm{x}^{(k)})&=\begin{bmatrix}\bm{\Sigma}_{p,0}^{(k)}&& & \bm{S}_{p}(\bm{x}^{(k)})\end{bmatrix}\in\mathbb{C}^{M\times kN} \\
		\bm{\alpha}_{p,1}^{(k)}&=\text{cat}\big\{\bm{\alpha}_{p,0}^{(k)},\bm{a}_{p}^{(k)}\big\}\in\mathbb{C}^{kN}
	\end{align}
\end{subequations}
are the augmented mode matrix and gain vector accounting for the target and the interference under $\mathcal{H}_{1}^{(k)}$. Accordingly, it can be verified that the decision rule based upon the \gls{gic} is\footnote{The reader may refer to~\cite{stoica-2004-model,stoica-2004-On} for details on the GIC rule.}
\begin{equation}\label{GIC-approx-k}
	\max_{\bm{x}^{(k)}\in \mathcal{G}^{(k)}}\mathcal{J}^{(k)}(\bm{x}^{(k)})\mathrel{\underset{\mathcal{H}^{(k)}_{0}}{\overset{\mathcal{H}^{(k)}_{1}}{\gtrless}}} 0
\end{equation}
where
\begin{align}
	\mathcal{J}^{(k)}(\bm{x}^{(k)})&=
	\sum\limits_{p=1}^{P} \bigg(\Big\| {\bm{\Pi}}_{p}^{(k)}(\bm{x}^{(k)})\bm{C}_{p}^{-1/2}{\bm{r}_{p}}\Big\|^2\notag
	\\ &\quad-\eta\, \text{rank}\Big\{{\bm{\Pi}}_{p}^{(k)}(\bm{x}^{(k)})\Big\}\bigg)\label{GIC-approx-k-J-2}\\ 
{\bm{\Pi}}_{p}^{(k)}(\bm{x}^{(k)})&=\Big(\bm{I}_{M}-{\bm{\Xi}}_{p}^{(k)}\Big)\bm{C}_{p}^{-1/2}\bm{S}_{p}(\bm{x}^{(k)})\notag
\\ &\quad \times \Big(\Big(\bm{I}_{M}-{\bm{\Xi}}_{p}^{(k)}\Big)\bm{C}_{p}^{-1/2}\bm{S}_{p}(\bm{x}^{(k)})\Big)^{+}\label{Scharf-decomposition-Pi}.
\end{align}
The following remarks are now in order. The matrix ${\bm{\Xi}}_{p}^{(k)}$ is the orthogonal projector on the interference subspace spanned by the columns of $\bm{C}_{p}^{-1/2}\bm{\Sigma}_{p,0}^{(k)}$, with the understanding that ${\bm{\Xi}}_{p}^{(k)}$ is the all-zero $M\times M$  matrix for $k=1$.  The matrix ${\bm{\Pi}}_{p}^{(k)}(\bm{x}^{(k)})$ is the orthogonal projector
on the part of the column space of $\bm{C}_{p}^{-1/2}\bm{S}_{p}(\bm{x}^{(k)})$ not contained in the interference subspace. In~\eqref{GIC-approx-k-J-2}, $\big\| {\bm{\Pi}}_{p}^{(k)}(\bm{x}^{(k)})\bm{C}_{p}^{-1/2}{\bm{r}_{p}}\big\|^2$  is the energy of the whitened measurement $\bm{C}_{p}^{-1/2}{\bm{r}_{p}}$ contained in the subspace spanned by the column of ${\bm{\Pi}}_{p}^{(k)}(\bm{x}^{(k)})$, while $\text{rank}\big\{{\bm{\Pi}}_{p}^{(k)}(\bm{x}^{(k)})\big\}$ is the model order under $\mathcal{H}^{(1)}_{k}$ and  $\eta$ is a penalty factor. Finally, the decision rule in~\eqref{GIC-approx-k} compares the maximum value of the scoring metric in~\eqref{GIC-approx-k-J-2} over all points in $\mathcal{G}^{(k)}$ with a threshold. When $\mathcal{H}_{1}^{(k)}$ is accepted, then the estimated location $\hat{\bm{x}}^{(k)}$ of the $k$-th detected target is the argument of the maximum.

The proposed decision logic sequentially solves the testing problems in~\eqref{GIC-approx-k} until no additional target is found  or $\mathcal{G}^{(k+1)}=\emptyset$  or
$k=K_{\max}$. If the procedure terminates at iteration $\hat{K}$, the number of detected targets is $\hat{K}-1$  if the detection threshold has not been crossed in the last test and $\hat{K}$ otherwise. We choose the penalty factor $\eta$ to obtain a desired probability of false alarm, namely, $\text{P}_{\text{fa}}=\text{Pr}\big(\max_{\hat{\bm{x}}^{(1)}\in \mathcal{G}^{(1)}}\mathcal{J}^{(1)}(\hat{\bm{x}}^{(1)})>0 \text{ under }  \mathcal{H}_{0}^{(1)}\big)$.  The overall procedure is referred to as the \gls{msdis}. 

\subsection{Mitigation of Small-Scale Localization Errors}
Small-scale localization errors due to the off-grid target placement may be detrimental in the implementation of the \gls{msdis}, as the estimated interference subspace at iteration $k\geq2$ may not fully contain the echoes of the $k-1$ previously-detected targets. We describe next a practical fix. 

Let $\tilde{\mathcal{G}}\supset \mathcal{G}$ be a finer grid of points uniformly-spaced in the inspected region, with inter-element spacing $\tilde{\Delta}_{g}\ll \Delta_{g}$, and let $
\mathcal{B}^{(i)}=\big\{\bm{g}\in \tilde{\mathcal{G}}:\; \big\|\bm{g}-\hat{\bm{x}}^{(i)}\big\|\leq c/W\big\}
$ be the set containing the points in $\tilde{\mathcal{G}}$ \emph{close} to $\hat{\bm{x}}^{(i)}$; finally, let
$\bm{E}_{p,n}^{(i)}$ be the $M\times |\mathcal{B}^{(i)}|$  matrix  whose columns are the signatures $\{\bm{s}_{p,n}(\bm{g})\}_{\bm{g}\in\mathcal{B}^{(i)}}$. After interference rejection, at the $p$-th receiver, we assume that the $n$-th echo of the target detected in the $i$-th iteration  belongs to the augmented subspace spanned by the columns of $\big(\bm{I}_{M}-{\bm{\Xi}}_{p}^{(i)}\big)\bm{C}_{p}^{-1/2}\bm{E}_{p,n}^{(i)}$. Since this matrix contains signatures corresponding to non-resolvable delays, it may be ill-conditioned and some care is required to avoid an unnecessary subspace enlargement. Let  $\lambda_{p,n,1}^{(i)},\ldots,\lambda_{p,n,B^{(i)}}^{(i)}$ be its squared singular values of arranged in a decreasing ordering and let $\bm{u}_{p,n,1}^{(i)},\ldots,\bm{u}_{p,n,B^{(i)}}^{(i)}$ be
the corresponding left singular vectors; then, we only retain the left singular vectors
corresponding to the ${U}_{p,n}^{(i)}$ largest singular values, where ${U}_{p,n}^{(i)}\geq 1$ is the smaller index such that
\begin{equation}	
	\frac{\big|	\hat{a}_{p,n}^{(k)}\big|^2}{|\mathcal{B}^{(i)}|}	\sum_{m=U_{p,n}^{(i)}+1}^{|\mathcal{B}^{(i)}|}\lambda_{p,n,m}^{(i)}< \epsilon
\end{equation}
and $\epsilon$ is a parameter ruling the significance of the singular values with respect to the noise floor. At this point,  the matrix  ${\bm{\Xi}}_{p}^{(k)}$ used in~\eqref{Scharf-decomposition-Pi} is replaced by the orthogonal projector on the augmented interference subspace spanned by 
\begin{equation}
	\Big\{\Big\{\bm{u}_{p,n,1}^{(i)},\ldots,\bm{u}_{p,n,U_{p,n}^{(i)}}^{(i)}\Big\}_{n=1}^{N}\Big\}_{i=1}^{k-1}.
\end{equation}

\section{Numerical Results}\label{sect:Simulation example}

\begin{figure}[!t]
	\centering
	\includegraphics[width=0.7\columnwidth]{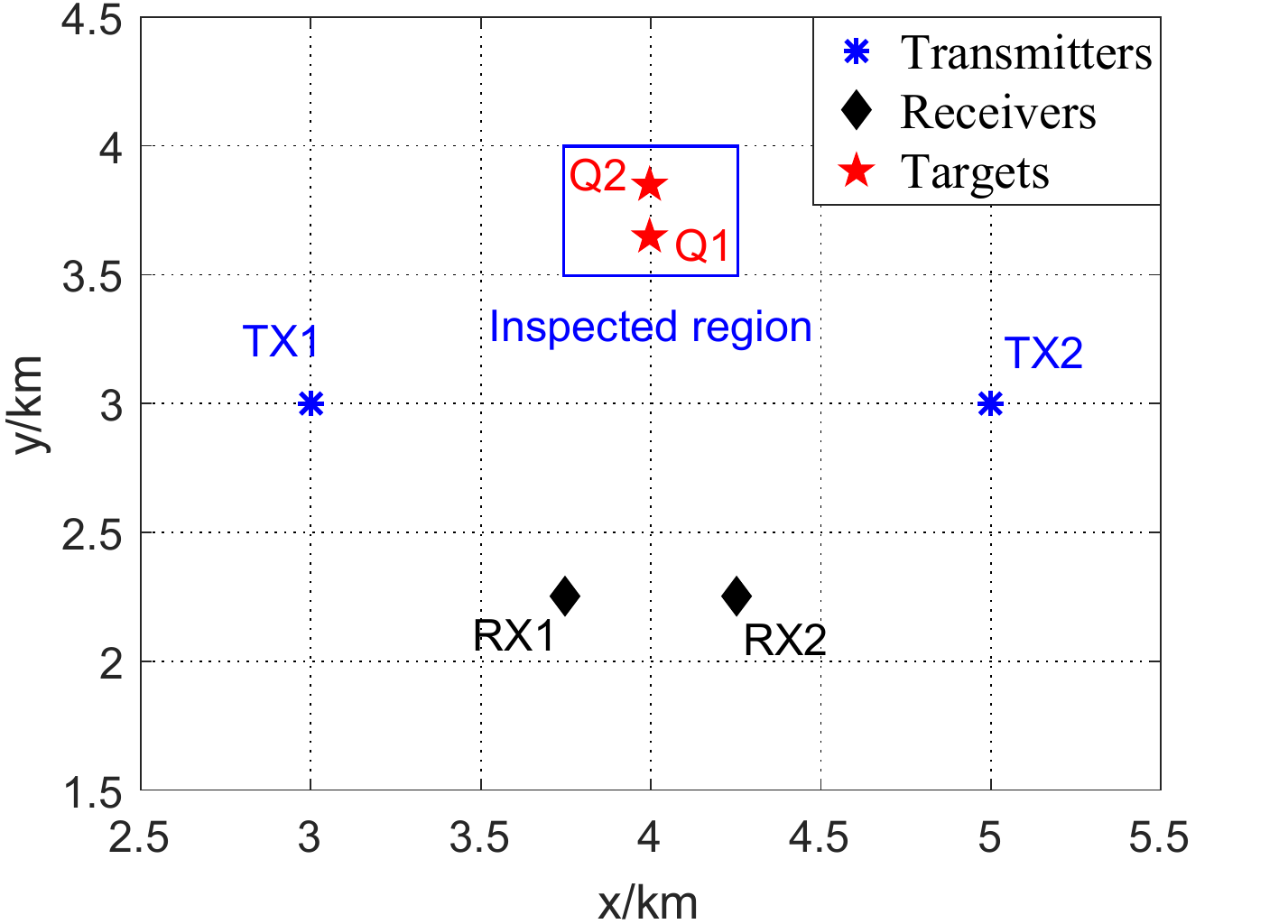}
	\caption{The positions of the transmitters, receivers and targets. }
	\vspace{-5pt}\label{fig: The positions of transmitters, receivers, and targets.}
\end{figure}

In the following example, we consider the $2\times 2$ MIMO radar in Fig.~\ref{fig: The positions of transmitters, receivers, and targets.} with two targets (namely, $\text{Q}_1$  and $\text{Q}_2$) located at $\bm{x}_{1}=(4000, 3650)~\text{m}$ and $\bm{x}_{2}=(4000, 3850)~\text{m}$. The \gls{snr} of the $k$-th target is defined as
\begin{equation}
	\text{SNR}_k=\frac{1}{NP}\sum_{n=1}^{N}\sum_{p=1}^{P}|a_{p,n,k}|^{2}\|\bm{C}_{w,p}^{-1/2}\bm{s}_{p,n}(\bm{x}_{k})\|^2.
\end{equation}  
Also, we consider the following time-coded radar waveforms 
\begin{equation}\label{eq:tx-pulse-train}
	\tilde{s}_{n}(t)=\sum_{\ell=1}^{64}b_{n,\ell}\psi\big(t-(\ell-1)T/64\big),\quad n=1,\ldots,N 
\end{equation} 
where $\psi(t)$ is a rectangular pulse with support in $[0,T/64]$ and $b_{n,1},\ldots, b_{n,64}$ is a random four-phase code sequences. Finally, $K_{\max}=5$, $W=10~\text{MHz}$, $\phi(t)=\sqrt{64/T}\psi(t)$, $T=6.4~\textmu\text{s}$, $T_{s}=0.05~\textmu\text{s}$, $\text{P}_{\text{fa}}=10^{-3}$,  $\Delta_{g}=10$~m, $\tilde{\Delta}_{g}=0.05$~m, and  $\text{SNR}_2/\text{SNR}_1=4$.  We assume that target $\text{Q}_1$ is the one of interest, and  the system performance is assessed in terms of its probability of detection ($\text{P}_{\text{d}}$) and the \gls{rmse} in the estimation of its position.  For comparison, we also include the performance obtained with the JDL algorithm with successive interference cancellation presented in~\cite{Yi-2020}  (shortly, JDL-SIC)  and with the \gls{glrtcd}, i.e., the GLRT-based detector ideally operating on a set of data where the echoes produced by the target $\text{Q}_{2}$ are not present, which represents the single-target benchmark.

\begin{figure*}[!t]  
	\centering
		\includegraphics[width=0.3\textwidth,draft=false]{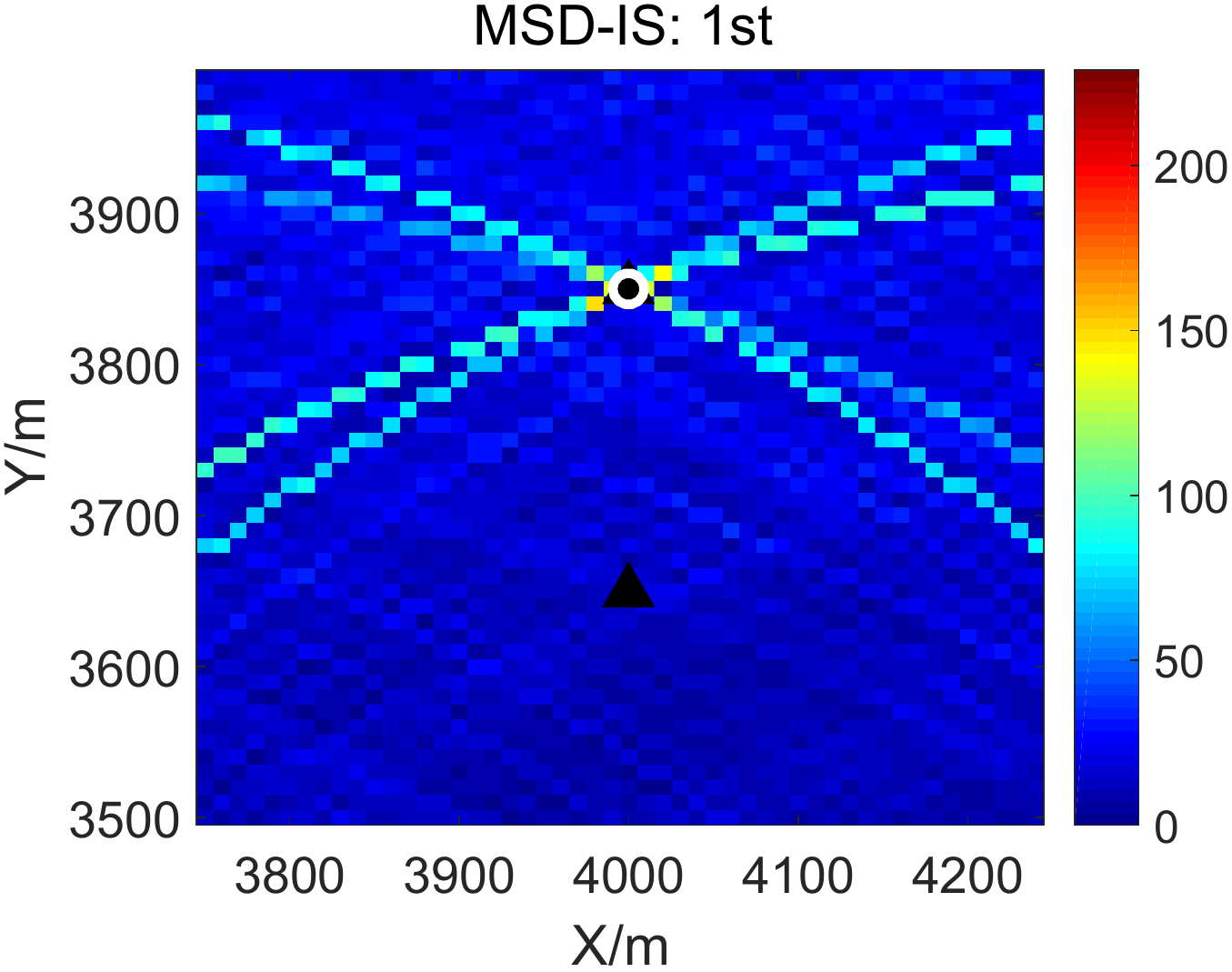}\quad\quad 
		\includegraphics[width=0.3\textwidth,draft=false]{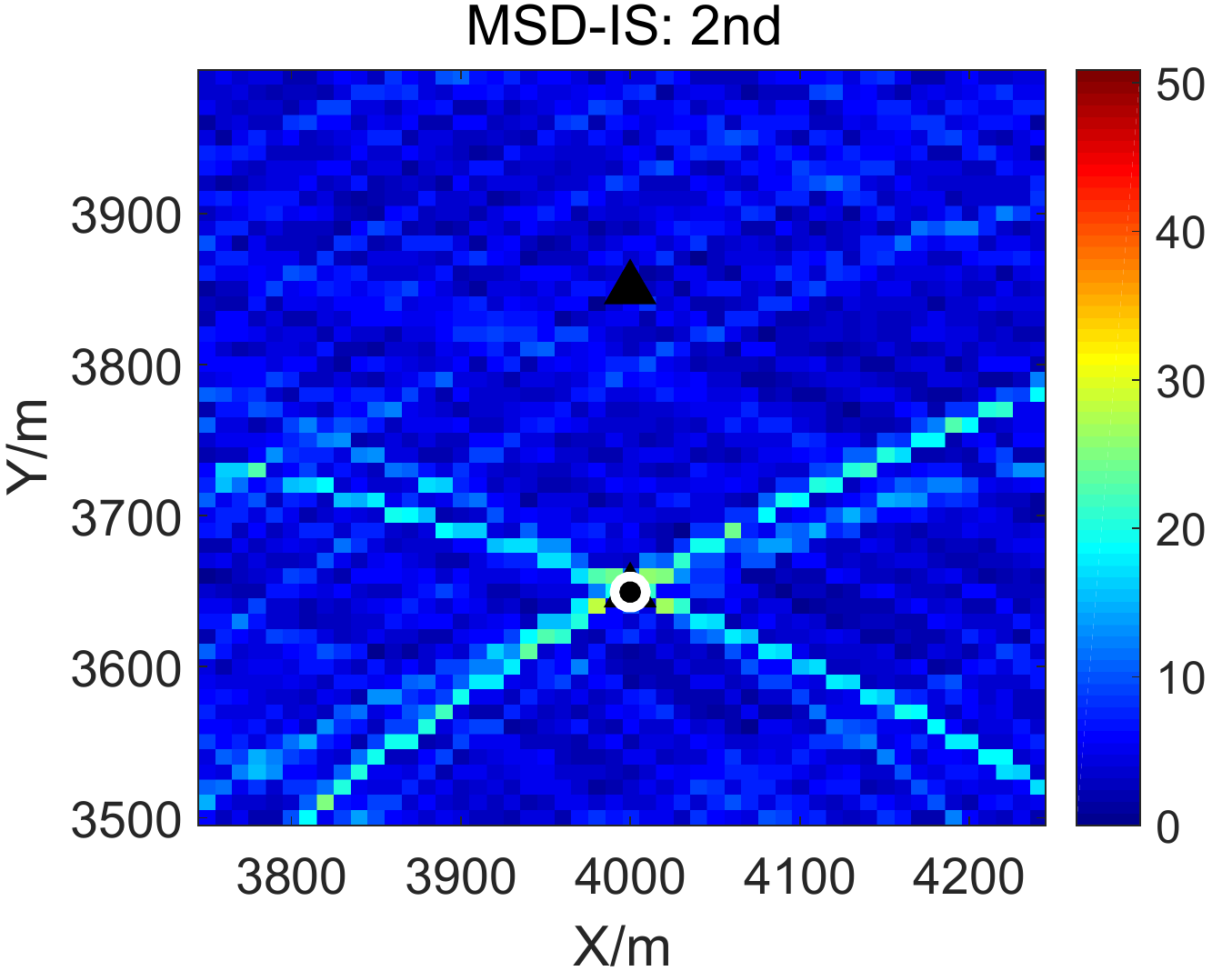}\quad\quad
		\includegraphics[width=0.3\textwidth,draft=false]{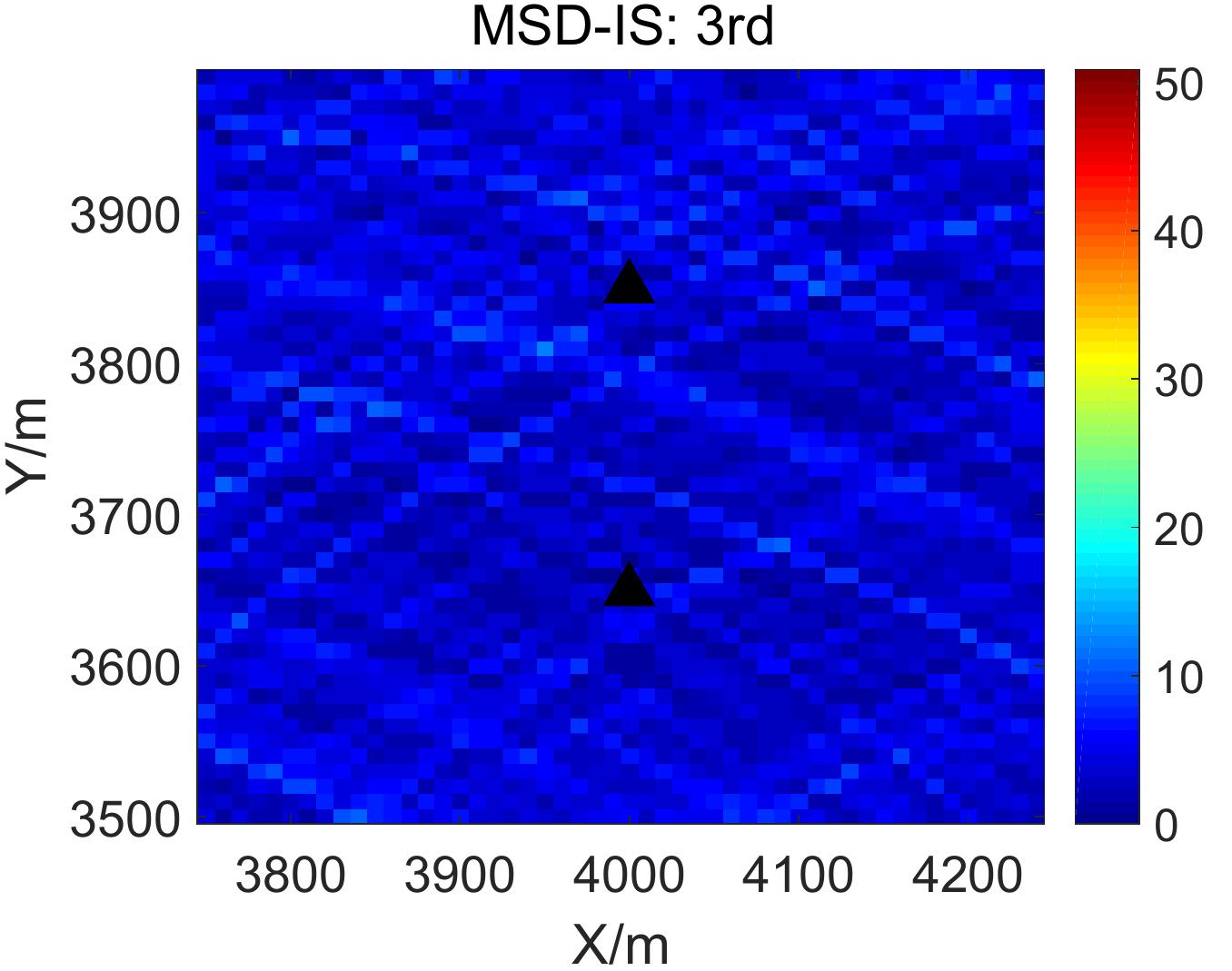}
	\\[10pt]
		\includegraphics[width=0.3\textwidth,draft=false]{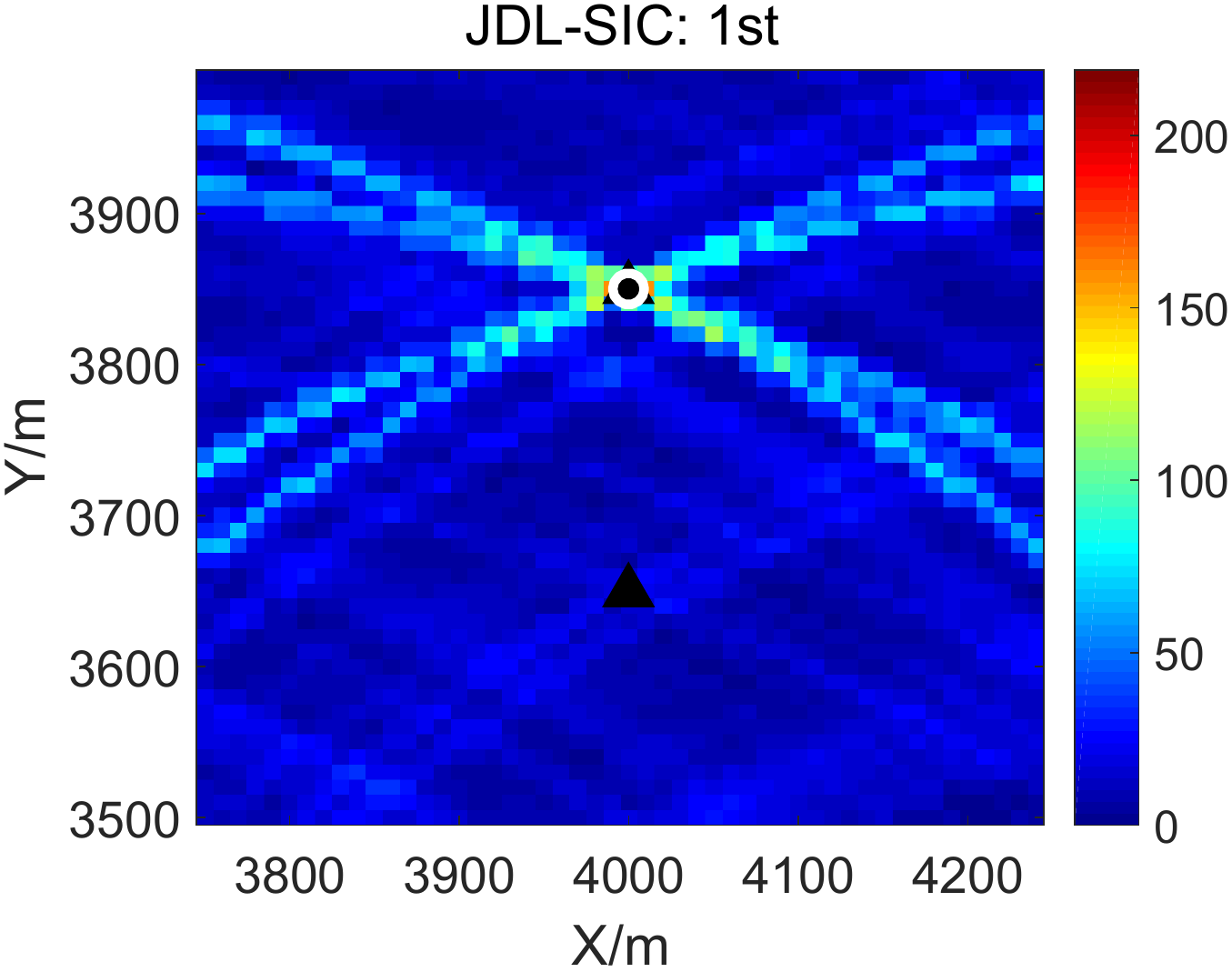}\quad\quad
		\includegraphics[width=0.3\textwidth,draft=false]{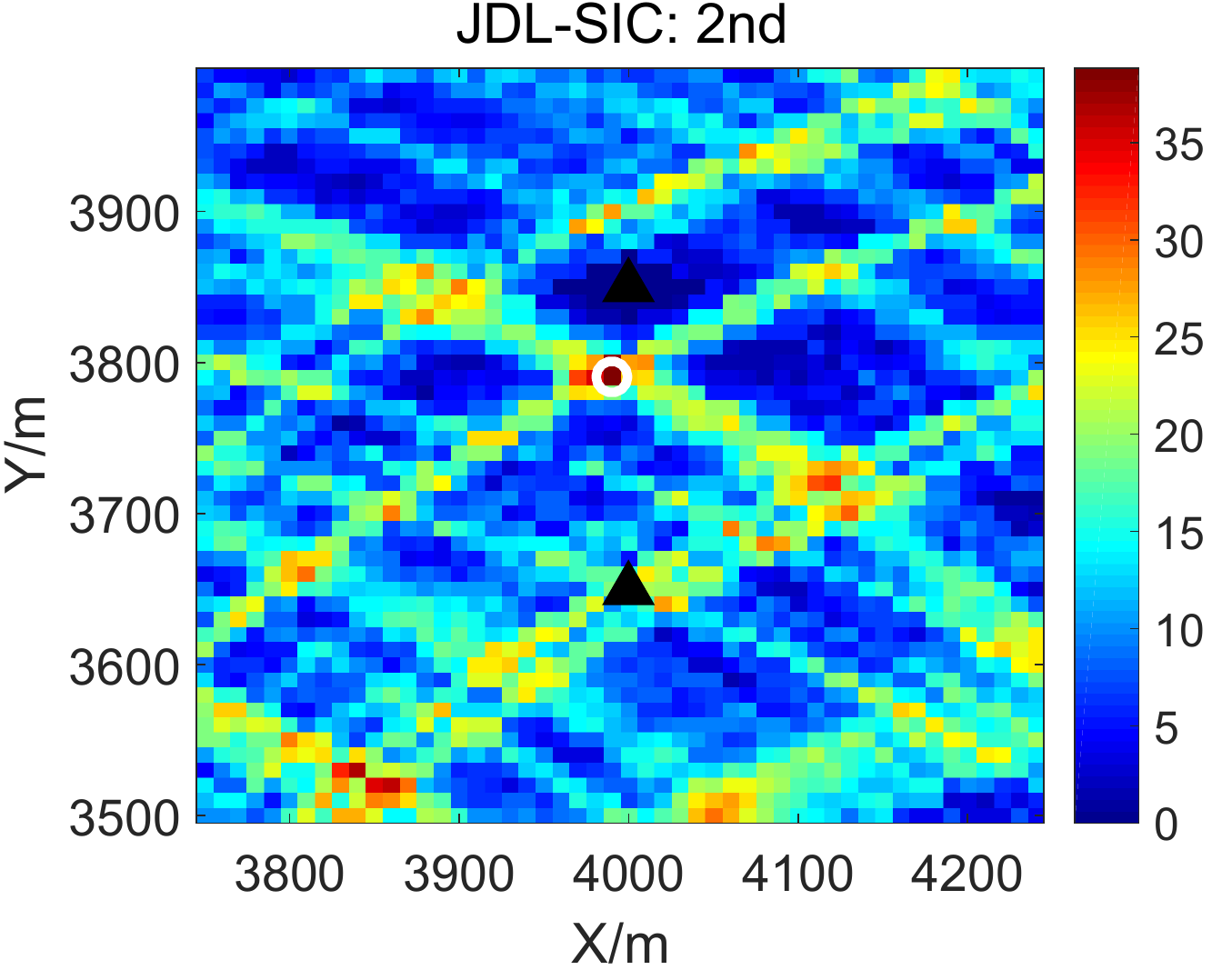}\quad\quad
		\includegraphics[width=0.3\textwidth,draft=false]{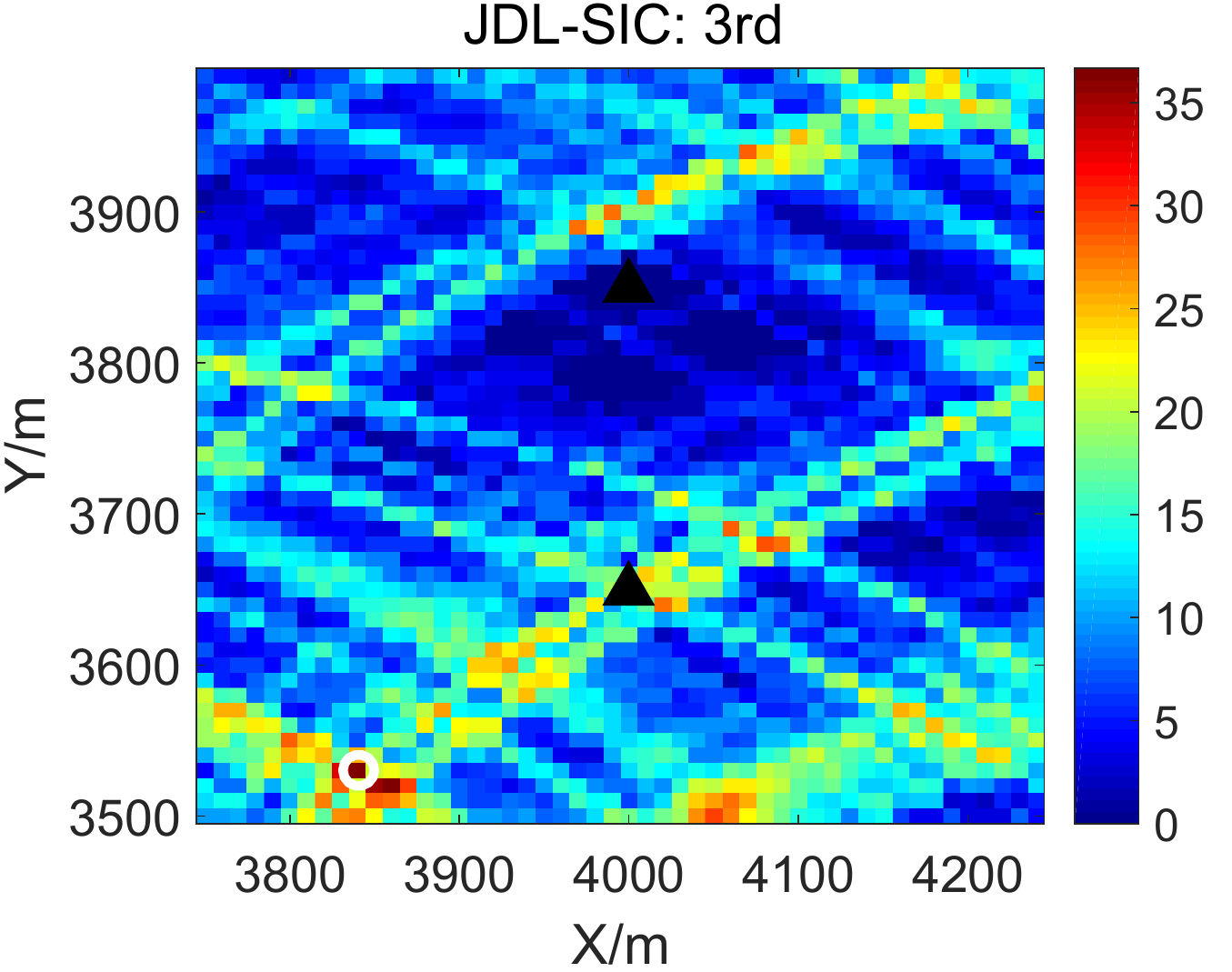}	
	\vspace{-5pt}\caption{ Output data planes of the \gls{msdis} (top) and JDL-SIC solutions (bottom) in three consecutive iterations (from left to right) when $\text{SNR}_{1}=13$~dB. The triangle markers are the true target positions; the circle marker is the estimated location of the target detected (if any) in each iteration.}
	\label{fig:MSD-IS/JDL-SIC, separable targets}
\end{figure*}

\begin{figure*}[!t]
	\centering
	\includegraphics[width=0.9\columnwidth]{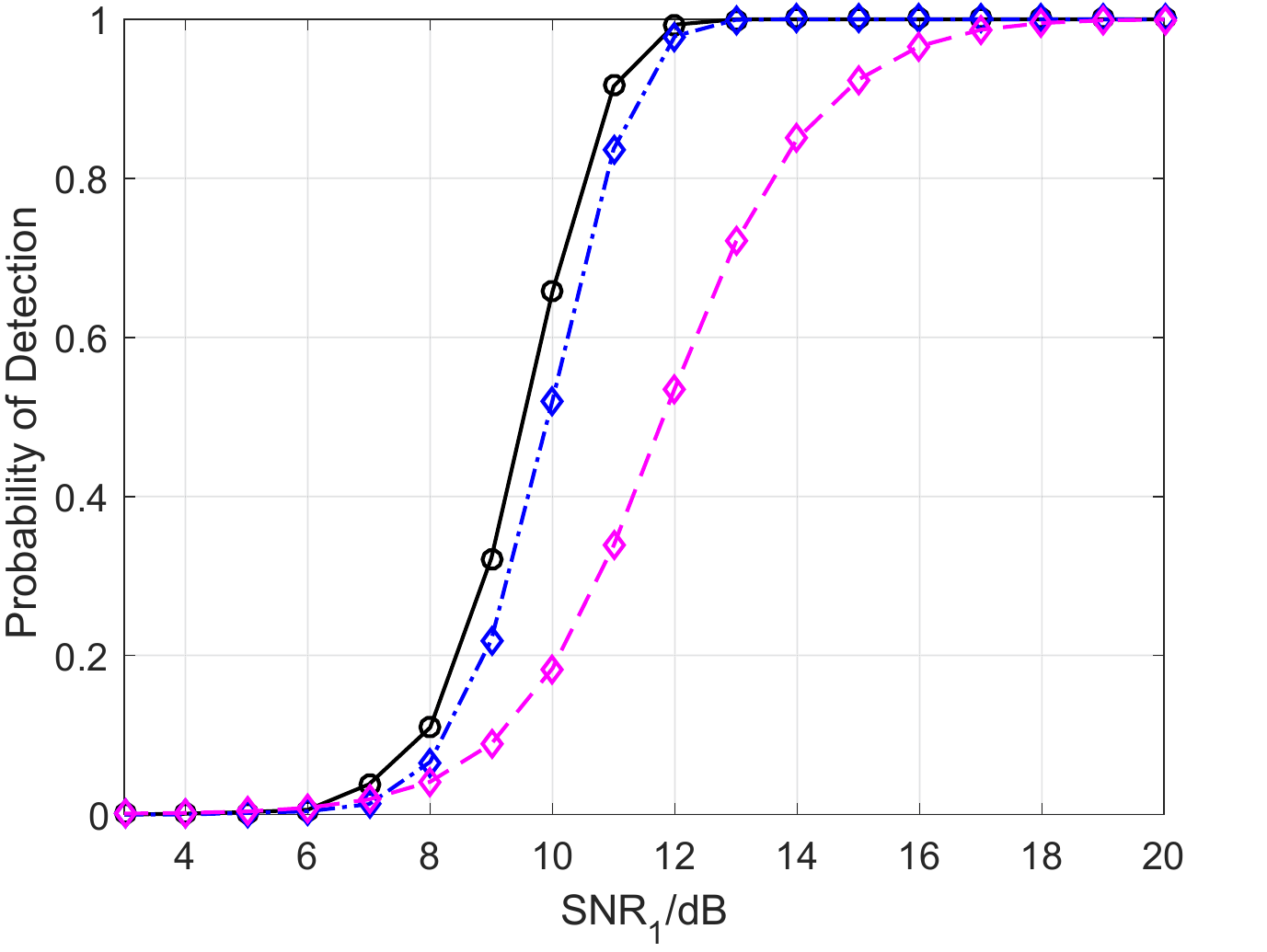}\quad\quad
	\includegraphics[width=0.9\columnwidth]{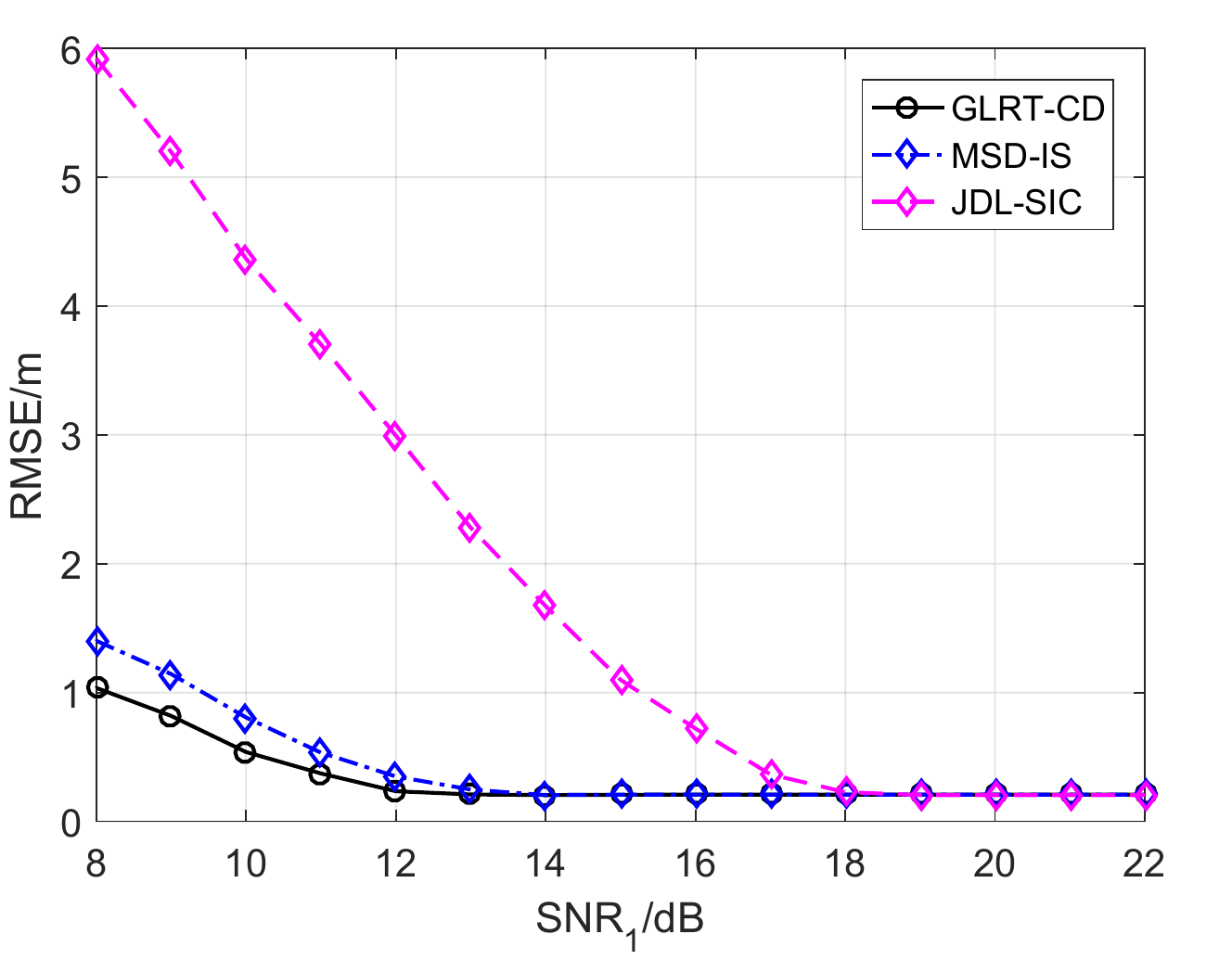}
	\vspace{-5pt}\caption{$\text{P}_{\text{d}}$ (left) and RMSE in the position estimate (right) of target $\text{Q}_1$ versus $\text{SNR}_{1}$.}
	\label{fig:partially_MC_1_Pd_RMSE_SNR}
\end{figure*}

We first compare the evolution of the \gls{msdis} and JDL-SIC solutions in a single snapshot. Fig.~\ref{fig:MSD-IS/JDL-SIC, separable targets} reports the output data planes (i.e., the value of the scoring metric in the inspected region) over three iterations. It is verified that the \gls{msdis} is able to correctly detect and localize one target at the time during the first two iterations and that no additional target is found in the third one; in particular, the output data plane is progressively cleaned by the mitigation of the interference caused by the previously-detected (stronger) targets. In contrast, the JDL-SIC only performs well in the first iteration (wherein the strongest target is found), but fails in the subsequent ones.  Finally, Fig. \ref{fig:partially_MC_1_Pd_RMSE_SNR} reports $\text{P}_{\text{d}}$ and RMSE in the position estimate of target $\text{Q}_1$ versus $\text{SNR}_{1}$.  It is seen that the \gls{msdis} performs close to the single-target benchmark, thus confirming its capability of detecting and localizing multiple targets even in the presence of a strong power imbalance among them.  On the other hand, the performance of the JDL-SIC is not satisfactory, since this receiver is not able to properly handle the residual cross- and auto-terms in the \gls{mf} output. Similar results have been obtained by including in the scene a larger number of targets, but the results have been omitted for brevity.

\section{Conclusions}\label{sect:Conclusion and next works}
This paper addresses the \gls{jdl} of multiple targets in a distributed \gls{mimo} radar when the adopted waveforms present non-ideal cross- and auto-correlation properties. After modeling the data collected by each receiver as the noisy superposition of an unknown number of subspace signals, we have proposed to iteratively extract one target at the time by considering a sequence of composite binary hypothesis testing problems. Each testing problem has been solved via a generalized information criterion, thus resulting into a subspace-based detector which mitigates the cross- and auto-terms of the previously-detected targets via a zero-forcing transformation.


\bibliographystyle{IEEEtran}
\bibliography{references}
\end{document}